\documentstyle[aps]{revtex}
\newcommand \ga{\raisebox{-.5ex}{$\stackrel{>}{\sim}$}}

\begin{document}
\draft
\date{Submitted to Physical Review B, 31 May 1995;
                      revision of 16 November 1995}

\title {Quantum saturation and condensation of excitons in Cu$_2$O:
a theoretical study}
\author{G.M. Kavoulakis, Gordon Baym and J.P. Wolfe}
\address {Department of Physics, University of Illinois at
 Urbana-Champaign\\ 1110 W. Green St., Urbana, IL 61801}
\maketitle

\begin{abstract}
\baselineskip=14pt

    Recent experiments on high density excitons in Cu$_2$O provide evidence
for degenerate quantum statistics and Bose-Einstein condensation of this
nearly ideal gas.  We model the time dependence of this bosonic system
including exciton decay mechanisms, energy exchange with phonons, and
interconversion between ortho (triplet-state) and para (singlet-state)
excitons, using parameters for the excitonic decay, the coupling to acoustic
and low-lying optical phonons, Auger recombination, and ortho-para
interconversion derived from experiment.  The single adjustable parameter in
our model is the optical-phonon cooling rate for Auger and laser-produced hot
excitons.  We show that the orthoexcitons move along the phase boundary
without crossing it (i.e., exhibit a ``quantum saturation''), as a consequence
of the balance of entropy changes due to cooling of excitons by phonons and
heating by the non-radiative Auger two-exciton recombination process.  The
Auger annihilation rate for para-para collisions is much smaller than that for
ortho-para and ortho-ortho collisions, explaining why, under the given
experimental conditions, the paraexcitons condense while the orthoexcitons
fail to do so.

\end{abstract}

\baselineskip=14pt
\section{Introduction}

    While Bose-Einstein condensation underlies the remarkable properties of
the strongly-interacting superfluid, liquid helium II, discovery of
Bose-Einstein condensation in nearly ideal gases has proven difficult.  Many
recent experiments have focussed on this challenge
\cite{Greytak,BEC93,Cornell}.  In the semiconductor Cu$_2$O, excitons
(electron-hole pairs bound by their Coulomb attraction) have exhibited
Bose-Einstein statistics \cite{HMB,SWM87} and indeed Bose-Einstein
condensation \cite{SWM90,LW}.  In these experiments, intense pulses of laser
light excite the crystal, creating a gas of (spin-triplet) orthoexcitons and
(spin-singlet) paraexcitons, split by an exchange energy $\Delta E \approx 12$
meV \cite{split,auger}.  The kinetic energy distribution of orthoexcitons as a
function of time from the onset of the laser pulse is observed by spectroscopy
of their photoluminescence (considerably more intense than that of the
paraexcitons).  In the classical (low density) regime, the energy distribution
is observed to be Maxwell-Boltzmann, described by an effective exciton
temperature.  In the quantum (high density) regime, the spectra exhibit
Bose-Einstein distributions, as shown in Figs.~1(a)-(c).  These spectra are
well fit in terms of an ideal gas with an instantaneous chemical potential,
$\mu$, and temperature, $T$.  From these parameters, one can directly
calculate the density of the gas.

    The densities obtained from the recombination spectra of Snoke et al. are
plotted in Fig.~1(d).  This figure shows the experimental results with a long
(10 ns) laser pulse (open circles) and with a short (100 ps) laser pulse
(black dots).  These experiments observe a ``quantum saturation'' of the
(spin-triplet) orthoexcitons, i.e., a tendency for them to move closely
parallel to the critical line without condensing \cite{SWM87}.  The critical
line, an adiabat (constant entropy per particle, $s=S/N$), has the form, $T_c =
(2\pi\hbar^2/mk_B)(n/g\zeta(3/2))^{2/3}$, where the degeneracy $g$= 3 for
orthoexcitons and 1 for paraexcitons.  We note that lines in the phase diagram
parallel (in a log-log plot) to the condensation line at higher temperature
correspond to adiabats, along which $\alpha \equiv -\mu/k_B T$ is constant,
since $s$ is only a function of $\alpha$ for an ideal Bose gas.

    Experimental measurements of the paraexcitons indicate condensation of
this component in unstressed \cite{SWM90} and uniaxially stressed crystals
\cite{LW}.  The stressed results are summarized in Figs.~2(a,b).  At high
densities the orthoexcitons decay very rapidly, due, as we shall see, to Auger
recombination and ortho-to-para conversion.  In contrast, the paraexcitons
decay very slowly even at high densities.  Figure~2(b) shows
density-temperature trajectories for both orthoexcitons (open circles) and
paraexcitons (black dots) for long pulses.  The stress splits the triply
degenerate orthoexciton level into three components and only the lower of the
three is significantly populated, leading to a closer proximity to the
condensation line than in the unstressed case.  No orthoexciton condensation
is observed.  The stress, however, does lead to clear evidence in the
paraexciton spectrum that the paraexcitons condense:  their energy
distribution is much sharper than that of the orthoexcitons at a given time.
Under these conditions the paraexciton density is determined from the relative
intensities of ortho and paraexcitons, combined with the spectroscopically
determined density of the orthoexcitons.  This analysis yields the striking
evidence for paraexciton condensation shown in Fig.~2(b).

    The non-stressed paraexciton spectrum also displays anomalies which are
interpreted as a result of condensation but are not so cleanly analyzed:  a
highly blue-shifted component is observed which may indicate a superfluid with
a large drift velocity.  The exact effect of the stress is not well
understood, but it is possible that it helps to localize better the two
components of the gas, making the analysis more straightforward \cite{bump}.

    In this study we identify the salient factors that lead to the quantum
saturation of orthoexcitons.  We find, by quantitative investigation of the
system, that the observed behavior can be understood in terms of
well-established kinetic processes of cooling, reheating, ortho-para
conversions, and recombination of the excitons.  The essence of the argument
is that the rate of change of the entropy of the excitons is a balance between
the entropy loss due to phonon cooling and heating following Auger
annihilation of exciton pairs into hot electrons and holes.  The rate of
phonon cooling, which is dominated by emission of acoustic phonons, varies as
$aT^{3/2}$ for exciton temperatures large compared with the ambient lattice
temperature.  On the other hand heating of the excitons from the Auger process
has, as we shall see, the form $bn_o$ proportional to the density of
orthoexcitons, $n_o$.  Thus the time rate of change of the orthoexciton
entropy per particle, $s_o$, is given in first approximation by
\begin{eqnarray}
T \frac{ds_o}{dt} = -aT^{3/2} +b n_o.
\label{orthoent}
\end{eqnarray}
Note that the right side vanishes along the adiabat given by $T^{3/2}/n_o
= b/a$, which is a stable fixed point of this equation, e.g., if the
temperature is too high, phonon cooling will increase, pulling the entropy
down to this adiabat, etc.  As we see in Sec.  VIc, where we present this
argument in greater detail, the stable adiabat is only slightly above the
condensation line.  Our qualitative and quantitatively detailed explanation of
the quantum saturation indicates that we have identified the critical
mechanisms of the exciton dynamics, enabling us to assert with confidence that
the paraexcitons do indeed condense under the experimental conditions.

    We model the time evolution of the gas from a rate equation approach.  In
Sec.~II we construct the rate equations that incorporate the relevant
dynamical processes, which we examine individually in Secs.~III-V.  In
Sec.~III we consider exciton-phonon interactions and show how the onset of
quantum statistics tends to suppress these cooling processes.  The effects of
ortho-para conversion mechanisms \cite{STW90,KM} are considered in Sec.~IV.
We find that the ortho-para conversion process is strongly modified for
excitons in the condensed phase.  We examine in Sec.~V the Auger-decay
process, in which two excitons annihilate to form an ionized electron-hole
pair \cite{SW90}, with the ionized electrons and holes subsequently
recombining into excitons.  This process tends to keep the gas away from the
degenerate region both by decreasing the exciton density and by indirectly
heating the excitons.  At high densities the orthoexcitons decay very rapidly
due to Auger recombination and ortho-to-para conversion; however, as we show,
the striking longevity of the paraexcitons, which is crucial to their ability
to cross the Bose-Einstein condensation boundary as seen in Fig.~2(b), is due
to the weakness of the para-para Auger process.  This weakness is a result of
the detailed band structure.  In Sec.  VI an analysis of the solutions to the
rate equations for a range of values of the input parameters allow us to see
the relative importance of the various physical processes in producing the
quantum saturation and condensation effects.  In the same section we discuss
the evolution of the orthoexciton entropy, and calculate the time for the
orthoexcitons to approach the adiabat of stability.  Finally in Sec.  VII we
summarize our conclusions.

\section{Rate equations}

    We treat the exciton gas as nearly ideal, and kept in thermal equilibrium
by the rapid exciton-exciton collisions, with a single temperature $T$,
generally greater than the lattice temperature $T_{\ell}$.
For example, for excitons scattering as hard spheres with a
scattering radius on the order of the exciton Bohr radius, the characteristic
scattering time is a few picoseconds for typical densities $10^{18}$-$10^{19}$
cm$^{-3}$.  The slower interconversion processes between the ortho and
paraexcitons do not allow them to come into chemical equilibrium quickly, so
that, in general, the chemical potential of the orthoexcitons $\mu_o$ differs
from that of the paraexcitons $\mu_p$.  We further assume that the lattice
maintains a constant temperature $T_{\ell}$ throughout the relaxation process.
At 30-50 ns after an intense 10 ns pulse the excitons assume a temperature
$\sim$ 5-6K, which we conclude is the lattice temperature in the excitation
region at that time, somewhat higher than that of the surrounding cold lattice
and the bath temperature (2K).

    In the experiments, photoexcitation by green laser light creates hot
electrons and holes within about a micrometer of the crystal surface.  These
hot carriers diffuse and within a nanosecond relax to form the excitons under
study.  During the first 5-20 ns the spatial extent of the excitonic gas under
consideration has been experimentally estimated to be about 30 $\mu$m.  We
make the simplifying assumption that the excitonic gas occupies a constant
volume over the entire time interval, which is valid at least for the first 7
ns.  Going beyond this approximation to include the full hydrodynamics of
expansion is a task for a future paper (see \cite{Link} for an initial
approach).  We also neglect inhomogeneities in the exciton gas and assume that
the ortho and paraexcitons occupy the same volume.  The parameters of the
problem, all time-dependent, are therefore the exciton temperature $T$, the
number of orthoexcitons $N_{o}$, and the number of paraexcitons $N_{p}$.

    In the normal (non-condensed) regime the chemical potentials $\mu_i$
($i=o,p$) of each of the species are functions of $N_i$ and $T$.  In the
condensed regime, where the chemical potential is zero, we take the number of
condensed particles $N_{i,cond}$ to be the equilibrium result,
$N_{i,cond}=N_i(1-(T/T_{ci})^{3/2})$, where $T_{ci}$ is the critical
temperature for condensation of the component $i$.

    Excitons are formed by the laser through creation of ionized electrons and
holes, with an excess kinetic energy of about 250 meV per pair, which then
combine in random spin states to give orthoexcitons and paraexcitons in a
three-to-one ratio.  The rate equations that describe how the total numbers
$N_{o}$ and $N_{p}$ of the two species change in time are then:
\begin{eqnarray}
 \frac {d N_{o}} {d t} = G_{o}(t) - \frac{N_{o}}{\tau_{lo}} - D N_{o,exc}
 + U N_{p,exc} - \frac{N_{o}}{\tau_{Ao}} +
 \frac{3}{4}\cdot\frac{1}{2} \left( \frac{N_{o}}{\tau_{Ao}} +
 \frac{N_{p}}{\tau_{Ap}}\right),
\label{21} \\
 \frac {d N_{p}} {d t} = G_{p}(t) - \frac {N_{p}} {\tau_{lp}} + D N_{o,exc}
 - U N_{p,exc} - \frac {N_{p}} {\tau_{Ap}} +
 \frac{1}{4}\cdot\frac {1}{2} \left( \frac {N_{o}} {\tau_{Ao}} +
 \frac {N_{p}} {\tau_{Ap}} \right),
\label{22}
\end{eqnarray}
where $G_i(t)$ is the laser production rate of excitons, and the
$\tau_{li}$ are the intrinsic radiative lifetimes ($\tau_{lo}\approx$ 300 ns
and $\tau_{lp}\approx$ 150 $\mu$s).  The quantity $D$ is the ortho to para
down-conversion rate, and $U$ is the para to ortho up-conversion rate; as we
show later only the non-condensed (non-zero momentum) excitons participate
significantly in the ortho-para interconversion processes, so that these terms
are effectively proportional to the number of non-condensed, or excited,
particles, $N_{i,exc}$.  The terms $-N_i/\tau_{Ai}$ describe the Auger
annihilation process, while the final terms account for the reformation of
excitons from the Auger-ionized electrons and holes.  The factor of 1/2 arises
because for each two excitons destroyed by the Auger process, only one
reforms, with probability 1/4 to be a paraexciton and 3/4 to be an
orthoexciton.  The right side of these equations is a function of the numbers
of excitons $N_{o}, N_{p}$ and their common temperature $T$.

    The third kinetic equation is the rate of change of the total internal
energy, $E=E_o+E_p$, of the gas.
\begin{eqnarray}
   \left(\frac{d E}{d t}\right)
   =  \left(\frac{\partial E}{\partial t}\right)_{phonon}
       +\left(\frac{\partial E}{\partial t}\right)_{o\to p}
       +\left(\frac{\partial E}{\partial t}\right)_{p\to o}
       +\left(\frac{\partial E}{\partial t}\right)_L
       +\left(\frac{\partial E}{\partial t}\right)_A
       +\left(\frac{\partial E}{\partial t}\right)_l.
\label{23}
\end{eqnarray}
The terms in this equation describe, respectively, the change in the
energy due to collisions of the excitons with the cold lattice (``phonon"),
phonon-assisted ortho-para interconversions, laser heating (L), heating from
Auger annihilations (A) and radiative recombination (l).  In fact, because the
radiative recombination lifetimes $\tau_{lo}$ and $\tau_{lp}$ are long, the
latter process is negligible in the three rate equations.  Since $E$ is a
function of $N_{o}, N_{p}$ and $T$, the rate equations can be equivalently
cast as equations for the rate of temperature change $\partial T/\partial t$,
and the exciton numbers.  Let us now examine each of the relaxation mechanisms
in turn.

\section{Phonon cooling of excitons}

    Excitons in Cu$_2$O interact with acoustic and optical phonons through a
deformation potential
interaction \cite{BS,Seeg}.  If the gas is above the condensation temperature
$T_c$, the rate of change of the energy of the exciton gas for the processes
in which an exciton emits or absorbs a phonon, shown in Figs.~3(a,b), is:
\begin{eqnarray}
  \left( \frac {\partial E} {\partial t} \right)_{phonon} =
  \frac{2\pi}{\hbar} \sum_{\vec k, \vec q}
   |M_{\vec q}|^2 \{n_{\vec k} (1+n_{\vec k + \vec q}) f_{\vec q} -
   n_{\vec k + \vec q}\,(1+n_{\vec k}) (1+f_{\vec q}) \}
   \hbar \omega_{\vec q}\,
   \delta(\varepsilon_{\vec k+\vec q}-\varepsilon_{\vec k} -
   \hbar\omega_{\vec q});
\label{31}
\end{eqnarray}
here $n_{\vec k}$ is the (ortho or para) exciton distribution, $f_{\vec
q}$ is the phonon distribution, $M_{\vec q}$ is the exciton-phonon interaction
matrix element, $\varepsilon_{\vec k}$ is the exciton energy and $\hbar
\omega_{\vec q}$ the phonon energy.  Below $T_c$, the ground state is occupied
by a macroscopic number of particles, so that we have an additional
contribution, Figs. 3(c,d), from condensed particles:
\begin{eqnarray}
  \left( \frac {\partial E} {\partial t} \right)_{phonon}^{cond} =
  \frac {2 \pi} {\hbar} \sum_{\vec q}
   {|M_{\vec q}|}^2 \{N_{cond} (1+n_{\vec q}) f_{\vec q}
   - (1+N_{cond}) n_{\vec q}\,
    (1+f_{\vec q})\,\} \hbar \omega_{\vec q}\,
   \delta (\varepsilon_{\vec q} - \hbar \omega_{\vec q}).
\label{32}
\end{eqnarray}
We define the exciton gas structure factor,
\begin{eqnarray}
   S(\vec q,\omega)=\sum_{\vec k} n_{\vec k} (1+n_{\vec k + \vec q})
   \delta(\varepsilon_{\vec k+\vec q} - \varepsilon_{\vec k} - \hbar \omega),
\label{33}
\end{eqnarray}
in terms of which Eq. (\ref{31}) takes the form:
\begin{eqnarray}
  \left(\frac{\partial E}{\partial t} \right)_{phonon} =
       \frac {2 \pi} \hbar \sum_{\vec q}
       |M_{\vec q}|^2 \hbar \omega_{\vec q}
       S(\vec q,\omega_{\vec q}) f_{\vec q}\,
       (1-e^{(\beta_{\ell} - \beta) \hbar \omega_{\vec q}}),
\label{34}
\end{eqnarray}
where $\beta_\ell = 1/k_B T_\ell$ and $\beta = 1/k_B T$.  In terms of
$v_{\pm}\equiv \beta[(\hbar^2/2m)(q/2 \pm {m \omega}/ {\hbar
q})^2-\mu]$, the structure factor becomes
\begin{eqnarray}
   S(\vec q,\omega)= \frac V {4 {\pi}^2} \frac {m^2} {\beta {\hbar}^4 q}
   \frac 1 {(1-e^{- \beta \hbar \omega})}
   \ln \left( \frac {1-e^{-v_+}} {1-e^{-v_-}} \right),
\label{35}
\end{eqnarray}
where $V$ is the volume of the exciton gas and $m$ is the exciton mass,
$\approx (3.0 \pm 0.2) m_e^0$ (where $m_e^0$ is the bare electron mass)
\cite{YS75,SBC91} for the orthoexcitons; in this analysis we assume that both
the ortho and the paraexcitons have the same mass,\footnote
{\baselineskip=14pt Our choice of the effective masses is based on the
experimentally known Rydberg of 97 meV for the excited exciton states, which
depends on the reduced mass of the electron and the hole and the dielectric
constant of the material.  A range of values for the effective electron and
hole masses have been reported \cite {Zhilich,Hodby}, and there are also
uncertainties in the dielectric constant \cite{OKeefe}.  The present
calculation is not critically sensitive to the choice of masses and dielectric
constant, so we choose the measured values $m_e = 0.84 m_e^0$, $m_h = 0.61
m_e^0$ and $\epsilon_0=7.11$ \cite{OKeefe} for the static dielectric constant,
which yield the excitonic Rydberg $\approx$ 99 meV, very close to the
experimental value obtained from the absorption spectrum of the excited
excitonic states.  The total orthoexciton mass is not equal to the sum of the
effective electron and hole masses $m_i$, likely due to the central cell
corrections \cite{YS75}, i.e., the corrections due to fact that the Bohr
radius is comparable with the lattice constant.  The central cell corrections
are also likely the reason that the binding energy of the 1s state
(approximately 153 meV) is much larger than the Rydberg for the excited
excitonic states.  This measured binding energy corresponds to a Bohr radius
$a=e^2/2 \epsilon_0 E_b \approx 7$\AA, where $E_b$ is the exciton binding
energy.}
and let $\varepsilon_{\vec k} = E_g - E_b + \Delta E + \hbar^2 k^2/2m$,
where $E_g$ is the energy gap energy ($\approx$ 2.17 eV in Cu$_2$O), $E_b$
is the binding energy, and $\Delta E$ is the exchange interaction (in which
an exciton virtually annihilates and reforms), which is non-zero for the
orthoexcitons and zero for the paraexcitons.

    For acoustic phonons $\omega_{\vec q}=v_{\ell} q$, where $v_{\ell}$ is the
longitudinal sound velocity.  We neglect the interaction of excitons with
acoustic transverse phonons, since it is about fifty times weaker than that
with the longitudinal phonons \cite{YS75,Tr}.  The square of the matrix element
for deformation potential scattering is \cite{Seeg}
\begin{eqnarray}
   |M_{\vec q}^{ac}|^2 = \frac {\hbar D_a^2 q} {2 \rho V v_{\ell}}
   \frac{1}{(1+(qa/4)^2)^4},
  \label{36}
\end{eqnarray}
where $D_a $ is the acoustic deformation potential, $\approx 1.8$ eV
experimentally \cite{SBC91}, $a$ is the exciton Bohr radius, $q$ is the phonon
wave vector, and $\rho \approx$ 6.0 gm/cm$^3$ is the mass density of the
material.  The quantity $(1+(qa/4)^2)^4$ in the denominator of Eq.  (\ref{36})
comes from the square of the Fourier transform of the relative 1s
electron-hole wave function, evaluated at $m_e = m_h$, where $m_i$ are the
effective electron and hole masses.  This correction term becomes significant
when the wavelength of the phonon is comparable to the Bohr radius, and is
actually negligible for thermalized excitons with $T \le 70$K, cf.
\cite{SBC91}.

    Combining Eqs.  (\ref{34}) and (\ref{36}), we find the rate of
acoustic-phonon cooling per exciton,
\begin{eqnarray}
   \frac{1}{N}\left( \frac {\partial E} {\partial t} \right)_{LA}
       = - \frac {(m  v_{\ell}^2)^2} {\hbar} \frac 1 {8 \pi^3}
        \frac m {m_{ion}} \frac{1}{n a^3} \left(\frac{a}{a_{\ell}}\right)^3
        \frac {(k_B T)^5 D_a^2} {(\hbar v_{\ell}/a_{\ell})^6 (m  v_{\ell}^2)}
\phantom{XXXXXXXXXXXXXXXXXXXXXX}
\nonumber \\ \phantom{XXXXXXXXXXXXXXXX} \times
    \int_0^\infty dx\, x^3 \left[ \frac{1}{e^x -1}
    -\frac{1}{e^{x T/T_{\ell}}-1} \right]
    \frac{1}{(1+(k_B T a x / 4 \hbar v_{\ell})^2)^4}
    \ln \left( \frac {1-e^{-y_+}}{1-e^{-y_-}} \right),
\label{37a}
\end{eqnarray}
where $n=N/V$ is the density of excitons, $a_{\ell} = 4.26${\AA} is the
lattice constant, $m_{ion} \equiv \rho a_{\ell}^3$ is a mean ion mass and
$a \approx 7$\AA\, for the ground state. The factors $v_\pm$ in Eq.
(\ref{35}) become $y_{\pm}=(m v_{\ell}^2/8 k_B T) [(x k_B T/m v_{\ell}^2) \pm
2]^2 -\beta\mu$.  The prefactor $(mv_{\ell}^2)^2/\hbar$ is $\approx$ 176
meV/ns. Equation (\ref{37a}) depends upon the quantum degeneracy factor,
$\alpha \equiv -\mu/k_B T$. Along lines of small constant $\alpha$, the
acoustic-phonon cooling rate per particle, with $D_a = 1.8$ eV from Snoke et
al., is
\begin{eqnarray}
      \frac 1 N \left( \frac {\partial E} {\partial t} \right)_{LA}
    \approx - 0.49 T^{3/2}(1-T_{\ell}/T) \text{ meV/ns},
\label{c5}
\end{eqnarray}
with $T$ measured in Kelvin; in the classical limit, the coefficient
becomes 0.62.

    Excitons also cool by emission of optical phonons, of which there are 15
modes at the zone center of Cu$_2$O \cite{Carabatos}.  The optical phonons
are:  $\Gamma_{25}^{-}$ (three-fold degenerate, of energy 11.4 meV),
$\Gamma_{12}^{-}$ (two-fold degenerate, of energy 13.8 meV), $\Gamma_{15}^{-}$
(three-fold degenerate, of energy 18.7 meV), $\Gamma_{2}^{-}$ (non-degenerate,
of energy 43 meV), $\Gamma_{15}^{+}$ (three-fold degenerate, of energy 64
meV), and $\Gamma_{15}^{-}$ (three-fold degenerate, of energy 79 meV).  They
group into two parts; the first eight with relatively lower energies, while
the remaining seven have relatively higher energies.  In calculating the
cooling rate of thermalized excitons we neglect the contribution of the second
group; even the lowest ones with $\hbar \omega = 43$ meV correspond to a
Maxwellian temperature of $2\hbar\omega/3k_B \approx 300$K, which suppresses
their contribution to exciton cooling.  In contrast, these high energy phonons
definitely need to be taken into account in calculating cooling of
non-thermalized high kinetic energy excitons, Sec.  V. All the lower energy
phonons have negative parity, as their group theory notation indicates.  Since
the parity of the excitons is even, the matrix element for the exciton
odd-parity optical phonon interaction vanishes at the zone center.  The
average squared matrix element for the deformation potential interaction with
an optical phonon of frequency $\omega_{i}$ can be written as \cite{Seeg}
\begin{eqnarray}
  |M_{\vec q}^{opt}|^2=\frac {{\hbar} D_i^2} {2 \rho V \omega_{i}},
\label{38}
\end{eqnarray}
where $D_i$ is the optical phonon deformation potential, averaged over the
whole Brillouin zone; for the $\Gamma_{15}^{-}$ phonon of zone-center energy
18.7 meV, $D_i \approx$ 0.17 eV/{\AA} in magnitude \cite{SBC91}.  As noted in
Ref.  \cite{SBC91}, the ratios of the deformation potentials of the low-energy
phonons can be set by comparison with the data of Yu and Shen \cite{YS75} for
the relative efficiency of the three ``three-phonon Raman lines.''  The
heights of these lines, which are simply due to hot luminescence following
emission of an optical phonon by an exciton, are proportional to the square of
the nonpolar optical deformation potentials involved, divided by the phonon
energy.  In the data of Ref.  \cite{YS75}, the $\Gamma_{15}^{-}$ phonon
emission line, which corresponds to the sum of the contributions of two TO
(transverse optical) and the LO (longitudinal optical) mode, has a maximum
intensity about three times higher than the two lower-energy phonon lines,
which are roughly equal in maximum efficiency.  Using this analysis Snoke et
al. have fixed the deformation potentials of the $\Gamma_{25}^{-}$ and the
$\Gamma_{12}^{-}$ phonons at 0.72 $\pm$ 0.08 times the $\Gamma_{15}^{-}$
deformation potential.

     The contribution from the optical phonons to the cooling rate per exciton
is,
\begin{eqnarray}
      \frac 1 N \left( \frac {\partial E} {\partial t} \right)_{opt}
   = - \sum_i \frac {(m  v_{\ell}^2)^2} {\hbar} \frac 1 {\pi^3}
     \frac m {m_{ion}} \frac 1 {n a^3}
     \frac {(k_B T)^2 (D_i a)^2}{(\hbar v_{\ell}/a_{\ell})^4}
     \left(\frac{1}{e^{\beta \hbar\omega_{i}}-1} -
    \frac{1}{e^{\beta_\ell\hbar\omega_{i}}-1} \right)
\nonumber \\ \times
   \int_0^{\infty} x\, dx
     \ln\left(\frac {1 - e^{-w_-}}{1 - e^{-w_+}} \right),
\label{310a}
\end{eqnarray}
where now the factors $v_\pm$ become $w_{\pm}\equiv (x \pm {\hbar
\omega_{i}}/{4 k_B T x})^2-\beta\mu$. The sum is over the
group of lower-energy optical phonons. Along lines of constant $\alpha$, and
small lattice temperature, the cooling rate per particle due to each of the
optical phonons varies as $T^{1/2} e^{-\beta \hbar \omega_{i}/2}$.

    From Eq.  (\ref{32}), the net additional energy loss per condensed
exciton, from excited excitons emitting an acoustic phonon and dropping into
the condensate is
\begin{eqnarray}
       \frac 1 {N_{cond}}
      \left( \frac {\partial E} {\partial t} \right)_{LA}^{cond}=
         - \frac {8(m v_{\ell}^2)^2} {\pi\hbar}
          \frac m {m_{ion}} \frac { (m v_{\ell}^2) D_a^2}
        {(\hbar v_{\ell}/a_{\ell})^3} \frac{1}{\left(1+(k_0 a/4)^2\right)^4}
       (n_{k_0}-f_{k_0}),
\label{312a}
\end{eqnarray}
where $k_0=2mv_{\ell}/\hbar$.  For emission of optical phonons,
\begin{eqnarray}
    \frac 1 {N_{cond}}
           \left( \frac {\partial E} {\partial t} \right)_{opt}^{cond}=
        - \sum_i \frac {(m v_{\ell}^2)^2} {2^{1/2} \pi\hbar}
        \frac m {m_{ion}} \left( \frac {a_{\ell}} a \right)^2
       \frac {(D_i a)^2 (\hbar \omega_{i})^{1/2}}
      {(m v_{\ell}^2)^{3/2} (\hbar v_{\ell}/a_{\ell})}
     (n_{k_i^{'}}-f_{k_i^{'}}),
\label{313a}
\end{eqnarray}
where $k_i^{'}= (2m \omega_{i}/\hbar)^{1/2}$.

    Combining these various rates, we find the rate of change of the internal
energy of the excitons per particle due to the exciton-phonon interaction
shown in Figs.~4(a,b) for acoustic and optical phonons, respectively.  In
these figures the temperatures of the exciton gas and the lattice are kept
fixed at 30K and 6K, respectively, as the chemical potential and therefore the
density is varied. As we see, for typical temperatures and densities,
cooling of excitons by acoustic phonons dominates cooling by optical phonons
by more than an order of magnitude.  Optical phonons do not contribute
significantly to the cooling since their energy is greater than that of the
majority of the excitons.  In contrast, optical phonons play the
dominant role in Auger heating, as we discuss in Sec. V, since the process
involves either free electrons and holes, or excitons of very large
kinetic energies.

    The results in Figs.~4(a,b) show a clear effect of degeneracy on the
exciton cooling.  The rate is constant when the exciton gas is in the
classical regime, but it decreases with increasing degeneracy, approaching a
lower value as $\alpha \to 0$.  This decrease occurs because energy and
momentum conservation limits the number of excitons that are able to interact
with phonons as the degeneracy increases.  As the gas approaches the
condensation phase boundary, the cooling mechanisms become less effective.
Furthermore the dependence of the acoustic phonon matrix element on momentum
transfer (Eq.  (\ref{36})) enhances this effect.  As we discuss in the last
section, the smaller cooling rate at high degeneracy plays a significant role
in the evolution of the system.  A related classical phenomenon has been
observed by Trauernicht et al.  \cite{Tr} in experiments on Cu$_2$O, where the
diffusion constant of paraexcitons shows a sharp rise at low temperatures due
to their average velocity falling below the sound velocity.

\section{Exciton interconversion processes}

    Orthoexcitons in Cu$_2$O are observed to convert rapidly to paraexcitons
\cite{KM} through an acoustic phonon-assisted process, depicted in
Figs.~5(a,b), dependent on temperature but independent of the gas density at
low densities \cite{STW90}.  At the zone center the orthoexcitons have
$\Gamma_{25}^{+}$ symmetry and the paraexcitons $\Gamma_{2}^{+}$.  Therefore,
conversion requires a $\Gamma_{25}^{+} \otimes \Gamma_{2}^{+}=
\Gamma_{15}^{+}$ phonon.  Since the acoustic phonons at the center of the zone
in Cu$_2$O have $\Gamma_{15}^{-}$ symmetry, the matrix element for the
conversion process is strongly suppressed.  Away from the zone center the
matrix element, calculated in the $\vec k \cdot \vec p$ approximation, has the
form \cite{Kane},
\begin{eqnarray}
      |M_c|^2 \sim k_o^2,
\label{41}
\end{eqnarray}
where $k_o$ is the orthoexciton wavevector.  In this approximation the
symmetry of the orthoexciton wavefunction is approximately that at the zone
center, plus a small component with $\Gamma_{15}^{-}$ symmetry of amplitude
$\propto k_o$.  We begin by calculating the down-conversion rate
(Figs.~5(a,b)) when the system is not in the condensed region.  We extend
the calculation to condensed excitons below.

    The down-conversion rate, $D$, is given by
\begin{eqnarray}
   D &\equiv &\frac{1}{N_{o,exc}}
  \left(\frac{\partial N_o}{\partial t}\right)_{o\to p}
      \nonumber\\
  &=&\frac{2\pi}{\hbar} \frac{1}{N_{o,exc}} \sum_{\vec k,\vec q}
  |M_c|^2 \{ n_{\vec k}^o (1+n_{\vec k - \vec q}^p\,)(1+f_{\vec q})
   \delta (\varepsilon_{\vec k}+\Delta -
   \varepsilon_{\vec k - \vec q} - \hbar \omega_{\vec q})\nonumber\\
    &&+n_{\vec k}^o f_{- \vec q}\, (1+ n_{\vec k - \vec q}^p\,)
    \delta (\varepsilon_{\vec k}+\Delta
    -\varepsilon_{\vec k - \vec q} + \hbar \omega_{\vec q}) \},
\label{42}
\end{eqnarray}
where $N_{o,exc}$ is the number of non-condensed orthoexcitons.  We
extract the matrix element here from experimental data for down-conversion in
the non-degenerate regime \cite{KM}, for which the classical limit of Eq.
(\ref{42}) applies.  We find for low lattice temperatures that in this limit,
$D \approx 0.014 T^{3/2} \text{ ns}^{-1}$ with $T$ measured in Kelvin.  To
determine the rate of energy change rate due to down-conversion we multiply
the first term in the summand of Eq.  (\ref{42}) by $(\Delta - \hbar
\omega_{\vec q})$ and the second by $(\Delta + \hbar \omega_{\vec q})$.

    The up-conversion rate, $U$, for the processes shown in Figs.~5(c,d), is
similarly,
\begin{eqnarray}
  U &\equiv& \frac 1 {N_{p,exc}}
  \left( \frac {\partial N_p} {\partial t} \right)_{p \to o}
  \nonumber\\
  &=&\frac {2 \pi}{\hbar}\frac{1}{N_{p,exc}} \sum_{\vec k,\vec q}
    |M_c|^2 \{ n_{\vec k + \vec q}^p (1+n_{\vec k}^o) (1+f_{\vec q})
    \delta (\varepsilon_{\vec k} +
     \Delta - \varepsilon_{\vec k - \vec q} + \hbar \omega_{\vec q})
   \nonumber\\
      &&+n_{\vec k + \vec q}^p\, f_{-\vec q}\, (1+ n_{\vec k}^o)
    \delta (\varepsilon_{\vec k} + \Delta -
    \varepsilon_{\vec k - \vec q} - \hbar \omega_{\vec q}) \},
\label{44}
\end{eqnarray}
where $N_{p,exc}$ is the number of non-condensed paraexcitons.  Since the
ortho-para energy splitting $\Delta E = 12$ meV is $\gg k_B T$ for temperatures
of interest, $U$ is much smaller than $D$.  To determine the rate of energy
change due to this process we multiply the first term of the summand of Eq.
(\ref{44}) by $-(\Delta + \hbar \omega_{\vec q})$ and the second by $(-\Delta
+ \hbar \omega_{\vec q})$.

    We turn now to the condensed excitons.  Equation (\ref{41}) implies that
down-conversion of condensed orthoexcitons ($k_o=0$) is suppressed by the
symmetry of the excitons and the phonons at the center of the zone of Cu$_2$O.
The up-conversion of condensed paraexcitons is not possible either, because
energy and momentum cannot be conserved for the parameters of Cu$_2$O.  Thus
only excitons in {\it excited} states participate in this mechanism.  The
conversion process primarily influences the orthoexcitons, since the
up-conversion rate of paraexcitons is small.  Condensation of orthoexcitons
would in fact help them to maintain their high density and lower the heating
by down-conversion.

    Figure~6 shows the calculated rates of the down and up-conversion
processes as a function of temperature at fixed $\alpha$.  As expected both
the down-conversion and the up-conversion rates increase with exciton
temperature.  On the other hand, both rates are suppressed as the degeneracy
increases.

\section{Auger process}

    The non-radiative direct and phonon-assisted Auger decay processes,
illustrated in Figs.~7,8, are the most important mechanisms for loss of
excitons at high densities.  Studies of this mechanism have been carried out
on excitons in Cu$_2$O \cite{auger,SW90}, and in electron-hole plasmas
\cite{GW81,BL58} where an electron recombines and excites either another
electron high in the conduction band, or a hole deep in the valence band.
Detailed calculations of the Auger rates are given in Ref.  \cite{auger}; we
summarize the results here.

    The band structure of the material plays an essential role in the Auger
annihilation of excitons.  Because the conduction and the valence bands have
the same (even) parity \cite{Dahl}, the rate of the direct Auger process
(Fig.~7) without Umklapp is negligible, and the Umklapp terms are small.
Although the phonon-assisted Auger process (Fig.~8) is suppressed by the
additional matrix element for phonon emission, the process is dominated by a
large non-Umklapp term when the phonon has negative parity.  The process is in
fact faster than the direct.  The rate of phonon-mediated para-para Auger
annihilations is negligible compared with the rates for ortho-ortho and
ortho-para; the reason is that the intermediate negative-parity band which
enters the process is very close in energy to the conduction band for the
orthoexciton recombination vertex entering ortho-ortho or ortho-para
collisions, but it is a deep valence band for the paraexciton vertex entering
para-para collisions.

    The Auger decay times appearing in Eqs.  (\ref{21}) and (\ref{22}) are
given by \cite{auger}
\begin{eqnarray}
  \frac{1}{\tau_{Ao}^{ph}}&=& \frac{C}{2} \left(n_o +\frac{1}{2}
        n_p\right),
\label{53} \\
  \frac{1}{\tau_{Ap}^{ph}}&=& \frac{C}{4} n_o,
\label{54}
\end{eqnarray}
for the phonon-assisted process, where the constant $C$ is of order 0.4
ns$^{-1}$, when the exciton densities, $n_i$, are measured in units of
$10^{18} \text{cm}^{-3}$.  The Auger decay rate depends only on the density of
excitons for low lattice temperatures.  Since the Auger matrix element is
constant when the thermal energy of excitons (a few meV) is much less than the
gap energy $E_g$ (=2.17 eV in Cu$_2$O), the decay rate depends on neither the
exciton temperature nor the statistics.  Auger collisions are not influenced
by the thermal motion of the excitons, and hence, the decay rate is the same
for degenerate and classical gases of the same density.

    The Auger process produces hot carriers within the exciton gas -- ionized
electrons and holes -- excited by an energy of order the energy gap.  The hot
carriers primarily lose this energy by emission of a cascade of phonons
(Fig.~9(a)) -- and to a lesser extent by scattering with excitons (Fig.~9(b))
-- and eventually reform excitons \cite{EMW93}.  To calculate the fraction of
Auger energy which heats the exciton gas we split the problem into two parts,
before and after the exciton is formed.  We shall see that only a small
fraction of the hot-carrier energy goes into heating the gas, because the
carriers are quickly cooled by rapid Fr\"ohlich emission of optical phonons.
In contrast the hot excitons have a slower phonon emission rate, and their
kinetic energy is more readily shared with the thermalized excitons.

    The cooling of the carriers by optical phonon emission (Fig.~9(a)) is
given by
\begin{eqnarray}
   \left( \frac {\partial E_{\vec k}} {\partial t} \right)_{opt}
   = -\frac {2 \pi} \hbar \sum_{\vec q} |M_{\vec q}|^2 \hbar\omega_{i}
   \delta(E_{\vec k} - E_{\vec k - \vec q} - \hbar\omega_{i}).
\label{57}
\end{eqnarray}
At high energies the Fr\"ohlich polar interaction \cite{Seeg} is the
dominant process for the emission and absorption of free carriers by optical
phonons.  If there is only one optical phonon branch, the matrix element for
the Fr\"ohlich interaction is,
\begin{eqnarray}
   |M_{\vec q}\,|^2 = \frac {2 \pi e^2} V
   \left(\frac{1}{\epsilon_{\infty}} - \frac{1}{\epsilon_0} \right)
   \frac{\hbar\omega_{i}}{q^2}.
\label{58}
\end{eqnarray}
Here $\epsilon_{\infty} = 6.46$
is the high frequency and $\epsilon_0(=7.11)$ is the static dielectric
constant; these two dielectric constants differ only slightly as a consequence
of Cu$_2$O being almost non-polar \cite{OKeefe}.
 Since Cu$_2$O has a large number of optical
phonon modes, Eq. (\ref{58}) can only be applied approximately;
for the lowest optical phonon of energy 11.4 meV, it leads to a rate
$\sim$ 50 eV/ns for the highest energy of the carriers ($\approx$ 1 eV); the
net rate should be even higher.

    The rate of energy change of a hot carrier of momentum $\vec k$ and energy
$E_{\vec k}$ due to collisions with thermalized excitons of momentum $\vec p$
and energy $\varepsilon_{\vec p}$ (Fig.~9(b)) is given by
\begin{eqnarray}
   \left( \frac {\partial E_{\vec k}} {\partial t} \right)_{exc}
    = -\frac {2 \pi} \hbar
   \sum_{\vec p,\vec q} |M_{\vec q}^{ce}|^2 n_{\vec p}\,(1+n_{\vec p+\vec q})
    (E_{\vec k} - E_{\vec k - \vec q}) \delta(\varepsilon_{\vec p + \vec q}+
    E_{\vec k - \vec q}-\varepsilon_{\vec p}-E_{\vec k}),
\label{55}
\end{eqnarray}
where $M_{\vec q}^{ce}$ is the carrier-exciton matrix element.  If we
assume a momentum transfer $q\gg p$, then
\begin{eqnarray}
 \left( \frac {\partial E_{\vec k}} {\partial t} \right)_{exc}
 \approx -\frac {2 \pi} \hbar \sum_{\vec p} n_{\vec p}
 \sum_{\vec q} |M_{\vec q}^{ce}|^2 \varepsilon_{\vec q}\,
 \delta(\varepsilon_{\vec q}+E_{\vec k - \vec q}-E_{\vec k}).
\label{56}
\end{eqnarray}
As we derive in the Appendix, the matrix element $M_{\vec q}^{ce}$ is
\begin{eqnarray}
   M_{\vec q}^{ce} \approx \frac{2\pi e^2 a^2}{\epsilon_0 V}
   \frac {(\lambda_e-\lambda_h)}{\left(1+(qa/4)^2\right)^3},
\label{56A}
\end{eqnarray}
where\footnote{\baselineskip=14pt As mentioned earlier the exciton mass is
not equal to the sum of the effective electron and hole masses; we assume,
however, that the ratios $m_e/m$ and $m_h/m$ are the same as those following
from the assumption $m=m_e + m_h$.} $\lambda_e = m_e/m = 0.58$ and
$\lambda_h = m_h/m = 0.42$.  This process gives a cooling rate $\sim$ 1 eV/ns,
some 2\% of the lowest optical phonon cooling rate, Eq.  (\ref{57}).
Furthermore, the ratio of the rates of carrier-exciton to phonon cooling
decreases with decreasing carrier energy.

    When the hot carriers lose sufficient energy they reform hot excitons
which heat the exciton gas by scattering with thermal excitons.  It is this
heating of the exciton gas which stands in the way of the excitons crossing
the phase boundary into the condensed region.  Cooling of hot excitons by
phonon emission is competitive, but does not heat the exciton gas.  Heating of
the excitons per Auger pair recombination is at most of order the exciton
binding energy.  To calculate the kinetic energy transfered to the exciton gas
by this Auger annihilation process, we note that the newly formed excitons
scatter against the thermalized excitons (Fig.~9(c)) and cool at a rate
\begin{eqnarray}
   \left( \frac {\partial \epsilon_h} {\partial t} \right)_{exc}
    = -\frac{\epsilon_h}{2} nv_{rel}\sigma(v_{rel})
    \equiv -\frac {\epsilon_h } {2\tau_{sc}},
\label{513}
\end{eqnarray}
where $\epsilon_h$ is the kinetic energy of a hot exciton,
$\sigma(v_{rel})$ is the exciton-exciton scattering cross section, $v_{rel}$
the relative velocity of the excitons, and $n$ the thermalized exciton
density.  At low energy the cross-section is $\sigma(v_{rel})= {4\pi (2
a_{sc})^2}$, with a single-exciton radius for exciton-exciton collisions,
$a_{sc}$.  We use $a_{sc} = 3.0$\AA\, which approximates the numerically
calculated interaction potential \cite{Haug} quite well.  At high energy the
cross section calculated in Born approximation taking into account Coulomb
interactions between the electrons and holes in the different excitons is
\begin{eqnarray}
  \sigma(v_{rel}) = \frac {33} {35} \pi a^2
 \left( \frac {e^2} {\hbar v_{rel} \epsilon_0} \right)^2.
\label{5513}
\end{eqnarray}
A simple interpolation between these limits is
\begin{eqnarray}
  \sigma(v_{rel})=\frac {4 \pi (2 a_{sc})^2}
   {1+(140/33)(k_{rel}a_{sc})^2},
\label{515}
\end{eqnarray}
where $k_{rel} = (m v_{rel}/ \hbar)$.
In the limit that the hot exciton energy is much larger than that of the
thermalized excitons, Eqs.  (\ref{513}-\ref{515}) give
\begin{eqnarray}
      \left( \frac {\partial \epsilon_h} {\partial t} \right)_{exc} =
     -\hbar \omega_{i}^2
    \left( \frac {a_{sc}} a \right)^2 n a^3
   \frac{2^{7/2} \epsilon_h^{3/2}}
  {(\hbar \omega_{i})^2} \left( \frac {\hbar^2} {m a_{\ell}^2} \right)^{1/2}
    \frac 1  {(1 + (280/33) m a_{sc}^2 \epsilon_h / \hbar^2)}.
\label{516}
\end{eqnarray}
The prefactor, $\hbar \omega_{i}^2$ is $\approx$ 198 eV/ns for the lowest
energy optical phonon of 11.4 meV.

    Hot excitons also cool by phonon emission (Fig.~9(d)) without heating the
exciton gas.  The rate of cooling of a hot exciton of momentum $\vec k$ and
energy $\varepsilon_{\vec k}$ by optical phonon emission is given by
\begin{eqnarray}
   \left( \frac {\partial \varepsilon_{\vec k}} {\partial t} \right)_{opt}
   = -\frac{2 \pi}{\hbar}\sum_{\vec q} |M_{\vec q}^{opt}|^2 \hbar\omega_{i}
   \delta(\varepsilon_{\vec k}-\varepsilon_{\vec k-\vec q}-\hbar\omega_{i}),
\label{58a}
\end{eqnarray}
with squared matrix element (\ref{38}).  Emission of optical phonons via
deformation potential coupling cools a hot exciton of kinetic energy
$\epsilon_h > \hbar\omega_{i}$ at a rate
\begin{eqnarray}
        \left(\frac{\partial \epsilon_h} {\partial t} \right)_{opt} =
     - \sum_i \hbar \omega_{i}^2
    \left( \frac {\epsilon_h} {\hbar \omega_{i}} - 1 \right)^{1/2}
      \frac m {m_{ion}}  \left( \frac {a_{\ell}} {a} \right)^2
   \frac {(D_o a)^2 {\epsilon_h}^{1/2}}
  {2^{1/2} \pi (\hbar \omega_{i})^{3/2} }
  \left( \frac {\hbar^2} {m a_{\ell}^2} \right)^{1/2}.
\label{510}
\end{eqnarray}
For hot excitons, all 15 optical phonon modes need to be taken into
account in the sum.  The measurements of Snoke et al. probe only the three
lowest optical phonons $\Gamma_{25}^{-}$, $\Gamma_{12}^{-}$ and
$\Gamma_{15}^{-}$, of energies 11.4, 13.8 and 18.7 meV, respectively.  For the
$\Gamma_{12}^{-}$ optical phonon of 13.8 meV, for example,
\begin{eqnarray}
      \left(\frac{\partial \epsilon_h} {\partial t} \right)_{\Gamma_{12}^{-}}
     \simeq
  - 0.16 \left( \frac {\epsilon_h}
   {\hbar \omega_{\Gamma_{12}^{-}}} -1 \right)^{1/2}
\text {eV/ns}.
\label{510a}
\end{eqnarray}
In addition, excitons of energy higher than $2 \hbar\omega_{i}$ can cool
by a fast parity-conserving two optical-phonon emission process \cite{SBC91}.
For the simultaneous emission of two $\Gamma_{12}^{-}$ phonons of 27.6 meV,
Snoke et al. have placed experimental limits on the cooling rate which
indicate that it is at least ten times larger than the single phonon rate.
They have also found that the two-$\Gamma_{25}^{-}$ phonon cooling mechanism
is too small to be observed.  The complete treatment of optical phonon cooling
including all phonon processes is beyond the scope of our present study and
indeed requires much more experimental input.  Nevertheless, we can use the
experimentally-derived exciton cooling rates of Snoke et al. involving the
three lowest optical phonon branches as a gauge of the total cooling rate for
all optical phonon processes.  We introduce the cooling parameter $\gamma
\,(>1)$ such that
\begin{eqnarray}
       \left( \frac {\partial \epsilon_h} {\partial t} \right)^{tot}_{opt}
 =   \gamma \left[ \left(\frac{\partial \epsilon_h}
			      {\partial t} \right)_{\Gamma_{12}^{-}} +
       \left(\frac{\partial \epsilon_h} {\partial t} \right)_{2\Gamma_{12}^{-}}
      \right].
\label{510b}
\end{eqnarray}
We shall see that a fit of the short-pulse experimental results yields
$\gamma\approx 4.5$.

     At energies lower than the lowest optical phonon energy only acoustic
phonons can be emitted; the cooling rate for coupling to longitudinal acoustic
phonons calculated with a deformation potential coupling is
\begin{eqnarray}
      \left(\frac{\partial \epsilon_h} {\partial t} \right)_{LA} =
   - \hbar \omega_{i}^2
   \frac m {m_{ion}} \left( \frac {a_{\ell}} {a} \right)^4
  \frac {2^{7/2} D_a^2}
 {3 \pi (\hbar \omega_{i})^2 \epsilon_h^{1/2}}
 \left( \frac {\hbar^2} {m a_{\ell}^2} \right)^{1/2}
\frac {z_0^2 (z_0 + 3)} {(1+z_0)^3} ,
\label{512}
\end{eqnarray}
where $z_0 = 2 \left[ (2 m \epsilon_h a^2)^{1/2} - (m v_{\ell} a)
\right]/ \hbar $.
In general the exciton-phonon interaction is not as effective at cooling
excitons as it is as for free carriers.  For the polar interaction this is due
to the electrical neutrality of the exciton and for the deformation potential
scattering the matrix element of the process involves the difference of the
deformation potentials for the conduction and the valence bands.

    The total cooling rate of a hot exciton due to phonon emission and
scattering with thermalized excitons is
\begin{eqnarray}
    \frac{d\epsilon_h}{dt} = \left(\frac{\partial \epsilon_h}{\partial t}
   \right)_{ph} +\left(\frac{\partial \epsilon_h}{\partial t}\right)_{exc},
\label{517}
\end{eqnarray}
where $(\partial \epsilon_h/\partial t)_{ph}$ is the total phonon cooling
rate of hot excitons, the sum of Eqs.  (\ref{510b}) and (\ref{512}).  The heat
transfered to the exciton gas, $E_A$, per Auger annihilation process is then
\begin{eqnarray}
   E_A = - \int_0^{t_0} dt \left(\frac{\partial \epsilon_h}{\partial t}
  \right)_{exc},
\label{519}
\end{eqnarray}
where $t_0$ is the time it takes for the hot exciton to fall to the mean
kinetic energy of a thermalized exciton.  This time must be found numerically
by solving for the time evolution of the kinetic energy of the hot exciton.

    We note that Auger heating depends strongly on the exciton density;
it depends only weakly on the exciton temperature because the higher the
kinetic energy of the exciton gas, the less energy is transfered from hot to
cold excitons. Auger heating does not depend on the degeneracy of the gas.
In the limit that the density of the thermalized excitons goes to zero, the
Auger heating goes to zero. In the limit that the exciton density is very
large, the Auger heating is equal to the binding energy $E_b$, since all the
energy goes to the excitons. The ratio $E_A/E_b$ as a function of the density
of the thermalized excitons is shown in Fig. 10, assuming that the exciton gas
temperature is kept fixed at $T=30$K.

The rate of energy change of the exciton gas due to Auger heating is
\begin{eqnarray}
      \frac{1}{N} \left( \frac {\partial E} {\partial t} \right)_A
   = \frac {E_A} {\tau_A^{ph}},
\label{520}
\end{eqnarray}
where $(\tau_A^{ph})^{-1} \equiv -(1/N)(\partial N/\partial t)_A = (C/4)
n_o$ is the total Auger scattering rate for the excitons, Eqs.
(\ref{53},\ref{54}).  Since the laser creates ionized electrons and holes with
an excess kinetic energy of about 250 meV per pair, an energy larger than
$E_b$, the heating of the exciton gas from hot carriers generated by the laser
is the same as the Auger heating $E_A$.  The rate at which the laser heats the
excitons is effectively $E_A$ times the laser production rate, and for a given
density and temperature of the exciton gas the total heating rate due to Auger
decay and the laser production is $E_A$ times the sum of the Auger rate and
the laser generation rate.

\section{Results}

    In the previous sections we have derived from experimental measurements
and theoretical considerations numerical values of the parameters entering the
rates of energy change of the exciton gas due to phonon cooling, the rates of
ortho-para interconversion and energy exchange, and the Auger decay rate.  The
processes of Auger and laser heating involve the uncertain cooling parameter
$\gamma$, Eq.  (\ref{510b}).  We now solve the kinetic equations
(\ref{21}-\ref{23}) for $N_o(t), N_p(t)$ and $T(t)$ numerically, assuming
reasonable initial conditions for these quantities at $t = 0$.  We choose
$\gamma$ to optimize agreement between our results and the experimental data,
for the case of short-pulse excitation, (subsection A; Fig. 11) and keep this
parameter fixed in the rest of our calculation.  We find that $\gamma \approx
4.5$, a reasonable number as an average of the cooling rate of a hot exciton
due to all the optical phonons of Cu$_2$O (including single and multi-phonon
processes) in terms of the $\Gamma_{12}^{-}$ plus $2\Gamma_{12}^{-}$
optical-phonon cooling rate.

    In the present calculations of the dynamics we assume that the
orthoexcitons are triply degenerate and, hence, we compare our orthoexciton
calculations with the zero-stress data.  No experimental determination of the
paraexciton density has been made in the zero-stress case.  We find, however,
that predictions for the paraexcitons at zero stress agree remarkably well
with the observed paraexciton behavior under stress.  This is not surprising
as the paraexciton multiplicity does not change with applied stress.

    The numerical simulations together with the argument, given in the
Introduction, that the orthoexcitons should approach an adiabat just above the
condensation line, put us in a position to identify the crucial factors that
lead to the quantum saturation of orthoexcitons, while allowing condensation
of paraxcitons.  At the outset, we note that no single process we consider
poses an obviously insurmountable barrier to Bose-Einstein condensation.  To
show in detail the physical effects of the various processes we take the
approach of first computing the behavior of the gas with all the known factors
in the equations, and then dissecting the problem by removing each of the
processes from the calculation in turn.

\subsection{``Short-pulse'' excitation}

    A mode-locked, cavity-dumped Argon-ion laser produces nanojoule pulses
with about a 100 ps length.  For the photoluminescence time resolution of
about 100 to 300 ps, this excitation pulse is effectively a delta-function in
time, and the evolution of the system is observed without creation by the laser
of further particles.  The excitons, whose energy relaxation and decay we are
dealing with, form on a timescale shorter than the detection limits; as we
shall see, most of the action occurs within a few nanoseconds. We compute
the behavior of both orthoexcitons and paraexcitons in this ``short-pulse''
case, even though the radiative efficiency of the paraexcitons is a factor
500 less than that of the (higher energy) orthoexcitons, making
difficult to observe the paraexcitons during this short time period.

    Figure 11 shows the result of our simulation with initial conditions, $n_o
=25\times10^{18}$ cm$^{-3}$ and $n_p = 8.33 \times 10^{18}$ cm$^{-3}$ at a
temperature of 60K, and $T_{\ell}$ fixed at 8K.  We assume in this and the
long-pulse case that the laser produces three orthoexcitons for each
paraexciton, as discussed earlier.  For comparison to experiment in this
figure and those that follow, we plot Snoke's orthoexciton data (diamonds) for
the long-pulse case between $n_o = 10^{18}$ cm$^{-3}$ and $2 \times 10^{19}$
cm$^{-3}$.  The solid and dashed curves represent the theoretical orthoexciton
and paraexciton trajectories, respectively, over the time t = 0 to 2 ns.  As
time progresses the excitons cool.  The dotted straight lines are the
Bose-Einstein condensation lines, the upper for paraexcitons ($g$ = 1) and the
lower for orthoexcitons ($g$ = 3).

    A principal result of our study is that, as in experiment, the
orthoexcitons move closely parallel to the phase boundary without crossing it,
following an adiabat, as a result of the balance between the phonon cooling
and the Auger heating, Eq.  (\ref{orthoent}).  We present the details of this
argument in subsection C. When the orthoexciton density falls sufficiently low
($<10^{18}$ cm$^{-3}$), and their temperature approaches that of the lattice,
the orthoexcitons move away from the condensation line; their density
continues to decrease due to ortho-to-para conversion and Auger recombination.

    The calculations further predict that the paraexcitons quickly cool below
their phase boundary and condense.  There is little decay in the density of
the paraexcitons over the 2 ns time interval; in fact, their density initially
increases due to the down-conversion of orthoexcitons.  Once the orthoexcitons
are gone, the paraexcitons decay on a scale of their recombination time, which
is longer than the chosen time interval.  Note that the two components cover
the same temperature range because we have assumed, as throughout, that they
are in good thermal contact with each other and thus have the same
temperature.

    Now let us examine how changing specific quantities in our calculation
changes the results, starting with the lattice temperature.  We expect that at
lower lattice temperatures, the orthoexcitons cool more efficiently.  We see
from Fig.~12(a) that this is true to a small extent.  With a lattice
temperature of 4K, the orthoexcitons still follow the same trajectory at high
density -- implying that the heating is associated with other processes -- but
as the gas density reduces below about $3 \times 10^{18}$ cm$^{-3}$ a brief
crossing of the phase boundary occurs.  Unfortunately, there is no direct
experimental control over the local lattice temperature:  the bath temperature
is already 2K and the rise in lattice temperature is indirectly caused by the
phonon emission of photoexcited carriers.  Matters could be worse:  we see in
the figure that a lattice temperature of 18K makes Bose-Einstein condensation
highly improbable for the orthoexcitons.

    The assumption of quantum statistics in the exciton-phonon cooling
processes has a marked effect on the quantum saturation, as shown in
Fig.~12(b).  Taking classical statistics (hypothetically) in the cooling
process, we see that the orthoexciton gas does not follow a constant-$\alpha$
trajectory at any stage.  The quantum effects shown in Fig. 4 -- reductions in
the cooling rates at high quantum degeneracy -- apparently have a deleterious
effect on the condensation.

    Now let us restore the quantum statistics and observe the effect turning
off ortho-to-para conversion.  As we see in Fig.~13 the paraexcitons, while
still predicted to condense, do not show an increase in density with time,
because one source of paraexcitons has been removed.  Still, Auger
recombination diminishes the density of both types of particles.  The
orthoexcitons, however, come considerably closer to their phase boundary
because one of their decay mechanisms has been deleted.  Yet, no condensation
is predicted for the orthoexcitons.  Ortho-to-para conversion is not
the dominant culprit in preventing their condensation.

    Next let us see how the Auger recombination process affects the system.
In the calculation shown in Fig.~14(a) we assume that the Auger rate (and,
therefore, the Auger heating) is reduced by 50\%.  This decrease
enhances the density of both species at a given time and lowers the
temperature of the exciton gas.  We see that the orthoexcitons are barely able
to condense in this case.  To carry this effect to the extreme we remove the
Auger recombination completely, with the result shown in Fig.~14(b).  In this
case the total number of excitons is nearly conserved on the 2 ns timescale
because the radiative lifetimes of the excitons are much longer.
Ortho-to-para conversion causes the density changes seen in the figure.  The
condensation of both the orthoexcitons and paraexcitons is now unavoidable.
Actually there are two reasons for this:  first, the decay of both species is
reduced, and, second, there is no Auger heating, which depends strongly on the
exciton density and pushes the excitons away from the phase boundary.

    Optimistically, we can say that Nature has been kind.  Were the Auger rate
50\% larger, it is unlikely that even the quantum saturation effect --
which, as we have seen here, indeed senses the quantum statistics -- would be
observable.

\subsection{``Long-pulse'' excitation}

    If the mode-locker is removed from the laser, the cavity-dumped mode
provides 10 ns-long pulses with about an order of magnitude more energy.  Now
we must incorporate the generation rate of the laser in the numerical
calculation.  We assume that the initial temperature of the exciton gas is
equal to the lattice temperature, which we take to be 6K.  To start the gas
off in the classical regime, we choose initial densities of $0.1 \times
10^{18}$ cm$^{-3}$ for the orthoexcitons and $0.033 \times 10^{18}$ cm$^{-3}$
for the paraexcitons.  We further assume that the temporal laser profile is
Gaussian and as before that three orthoexcitons are produced for each
paraexciton.

    The principal result of our calculation is shown in Fig.~15.  Now the
orthoexcitons initially move up the phase boundary, towards higher density and
temperature, and decay closely along the same path.  This is precisely the
remarkable quantum-saturation behavior which we hoped to reproduce
theoretically and understand.  In addition, we see that the paraexcitons heat
up with the orthoexcitons, but they are able to cross their phase boundary at
an early stage, implying that the gas has a condensed fraction.  At the peak
of the laser pulse, both orthoexciton and paraexciton densities begin to
decrease.  The paraexcitons retain a high density as they cool and eventually
reach a condensed fraction of over 90\%!  The numerical simulation is
remarkably similar to the data of Lin and Wolfe \cite{LW} in their uniaxially
stressed sample.  As previously stated, we do not expect major changes in the
calculated behavior for the stressed and unstressed cases.

    For clarity, we reproduce in Figs.~16 the results of Fig.~15, showing
separately the build-up and decay processes.  In our calculation, a pulse with
a full-width at half-maximum of 3 ns is used, as shown in the insets of
Figs.~16.  Figure~16(a) plots the (upwardly rising) gas trajectories for the
first half of the laser pulse, and Fig.~16(b) shows the (downward falling)
results for the second half of the laser pulse.  We note that the initial
conditions of the ``decay phase'' differ from those of the short-pulse
experiments:  here, the ``initial'' paraexciton density is considerably higher
than the orthoexciton density, because the ortho-to-para conversion process
has had the entire first half of the laser pulse to pump up the paraexciton
density.

    Again, we test the effect of quantum statistics in the phonon cooling
processes.  Figure~17 shows that in the case of ``classical'' phonon-cooling,
both orthoexcitons and paraexcitons are predicted to condense.  Thus, the
quantum statistics play a crucial role in keeping the orthoexcitons away from
the condensed region.

    In Fig. 18 we show the result of our calculation if we neglect the
ortho-para interconversion mechanism.  As seen from this figure, the
orthoexcitons condense, but still tend to move along the condensation line.  A
major difference between this graph and the data shown in Fig.~2(b) is the
faster decay rate here of the paraexcitons at late times.  The reason for this
is the lack of the ortho-to-para conversion mechanism, which balances the loss
of paraexcitons due to the Auger process.

    In Fig. 19 we show the result of our calculation if the Auger process
(annihilation and heating) is neglected.  Both species condense for the same
reasons as those given for the short-pulse excitation, in Fig. 14(b).  The
effect of the Auger {\it heating} itself (with unmodified Auger annihilation)
is shown in Figs.~20(a,b), where we assume that the scattering radius
$a_{sc}$ for carrier-carrier scattering is ten times smaller or larger,
respectively, than assumed in Fig.~15.  This scattering length influences the
kinetic energy transfered from the Auger-ionized carriers to the thermalized
excitons.  In the first case both species condense because of the low Auger
heating.  In the second case, however, the behavior of excitons hardly changes
from that of the initial calculation, Fig.~15.  The reason for this is that,
according to Eq.  (\ref{515}) the cross section for the scattering is
independent of the exciton radius at large radius, and in this limit the rate
of energy transfer is dominated by the scattering process.

    Finally in Fig. 21 we investigate the result of changing the Auger decay
constant $C$ and the acoustic phonon deformation potential $D_a$. The constant
$C$ determines the Auger decay rate and consequently the Auger heating rate.
Also $D_a$ determines the dominant phonon cooling rate, which
is the acoustic phonon mechanism. For values of $C$ and $D_a$ inside the
shaded region, the orthoexcitons move along and closely to the condensation
line, i.e., between the two critical lines for condensation of the ortho and
the paraexcitons.  The region corresponds to $0 \le \alpha \le 0.45$.  For
values of $C$ and $D_a$ above the shaded region, the orthoexcitons condense
because of the more effective phonon cooling, the low Auger decay rate and the
low Auger heating.  Below the shaded region the orthoexcitons move away from
the degenerate regime, towards the classical regime, because of the low
phonon-cooling rate, the high Auger decay rate and the high Auger heating.

\subsection{The entropy of orthoexcitons}

    In the Introduction we briefly outlined the argument that the balance
between the rates of phonon cooling and Auger heating causes the orthoexcitons
to approach a critical adiabat, a line of constant entropy per particle of the
orthoexcitons, $s_o$, lying slightly above the condensation phase boundary.
We now spell out the calculation in detail.  The rate of change of $s_o$ is
given by
\begin{eqnarray}
    T \frac {ds_o} {dt} = \frac 1 {N_o} \frac {dE_o} {dt}
   -\frac{Ts_o+\mu_o}{N_o}\frac{dN_o}{dt} +\frac{P_o}{N_o}\frac{dV}{dt},
\label{s1}
\end{eqnarray}
where $E_o$ is the total energy and $P_o$ the pressure of the orthoexciton
gas.

    The dominant contributions to the right side of Eq.  (\ref{s1}) are the
phonon cooling and Auger heating, which enter the first term.  The second term
is negligible, since particle numbers change relatively slowly.  For short
pulses both the ortho-to-para conversion and the Auger decay rates are
proportional to $n_o \sim T^{3/2}$ along the phase boundary; the total decay
rate is $\approx$ 1 ns$^{-1}$ for (orthoexciton) density $10^{18}$ cm$^{-3}$
\cite{SW90}.  For the case of long pulses the net production rate (laser
production minus decay processes) is even smaller.  For small values of the
chemical potential, therefore, the second term will be on the order of the
kinetic energy times the decay rate, which is more than an order of magnitude
smaller than the phonon cooling and the Auger heating rates.  In addition, the
final term is negligible for small expansion of the gas, which we have
assumed.

    The slow rate of change of particle numbers also allows us to relate the
first term on the right side of Eq.  (\ref{s1}) to the total change of the
energy of the excitons (ortho plus para) by
\begin{eqnarray}
   \frac 1 {N_o} \frac{dE_o}{dt}
  = \frac{N}{N_o}\frac{C_o(T)}{C_o(T)+C_p(T)} \frac 1{N}\frac{dE}{dt},
\label{s2}
\end{eqnarray}
where $C_{i}=(\partial E_i/\partial T)_{N_i,V}$ is the heat capacity of
the $i^{th}$ component at constant volume and number of particles.  For the
excitons close to the Bose-Einstein condensation phase boundary, $C_p/C_o
\approx N_p/N_o$, so that $(1/N_o)(dE_o/dt) \approx (1/N)(dE/dt)$.

    The acoustic-phonon cooling rate per exciton, for excitons with $\alpha$ =
0.15, is $\approx -0.56 T^{3/2} (1-T_{\ell}/T)$ meV/ns [cf.  Fig. 4a].  Since
we are interested in $T \gg T_{\ell}$, we neglect the term $T_{\ell}/T$.

    In the case of short-pulse excitation, the paraexciton density is
approximately constant, $\approx 10^{19}$ cm$^{-3}$, due to a balance between
the ortho-to-para conversion process and the Auger loss of paraexcitons.
Since the Auger heating of orthoexcitons per particle is a function of the
total exciton density ($\ga 10^{19}$ cm$^{-3}$), we assume that $E_A \approx
0.77 E_b$ [cf.  Fig.  (10)] and neglect its weak density dependence.  Then the
Auger heating rate per particle varies approximately as $11.8 n_o$ meV/ns with
$n_o$ measured in units of $10^{18}$ cm$^{-3}$.  For the case of long-pulse
excitation, the Auger heating is a bit smaller, closer to $E_A \approx 0.64
E_b$, since the total density is smaller in this case, but still the density
dependence of $E_A$ is weak for $n \ge 5 \times 10^{18}$ cm$^{-3}$.

    With these approximations Eq.  (\ref{s1}) takes the form
\begin{eqnarray}
     T\frac{ds_o}{dt} \approx -aT^{3/2}+bn_o,
\label{s3}
\end{eqnarray}
where, for $T$ in Kelvin and $n_o$ in units of $10^{18}$ cm$^{-3}$, $a
\approx 0.56$ meV/ns, and $b \approx 11.8$ meV/ns for short pulses and
$\approx 9.8$ meV/ns for long pulses.  Note that the right side of Eq.
(\ref{s3}) vanishes on the adiabat for which, in the same units, $n_o \approx
0.048 T^{3/2}$ for short pulses, and $n_o \approx 0.058 T^{3/2}$ for long
pulses, and that this equation predicts that the orthoexcitons approach this
adiabat as a stable fixed point.  The adiabat of stability, as calculated
approximately here, is shown as a dash-dot line in Fig. 11 for short
pulse-excitation and Fig. 15 for long-pulse excitation.

    To calculate the characteristic time, $\tau_*$, for the orthoexcitons to
approach the adiabat of stability, we regard Eq. (\ref{s3}) as an equation for
$\alpha$.  Near this adiabat, on which $\alpha \equiv \alpha_*$,
\begin{eqnarray}
   \frac{d\alpha}{dt} = -\frac{\alpha - \alpha_*}{\tau_*},
\label{s5}
\end{eqnarray}
where
\begin{eqnarray}
      \tau_* = -\frac{T}{b}\left(\frac{\partial s_o}{\partial n_o}\right)_T
    \approx \frac{5\zeta(5/2)}{2\zeta(3/2)} \frac{k_B}{aT^{1/2}};
\label{s6}
\end{eqnarray}
the latter form holds near the phase boundary.  Numerically $\tau_*
\approx 0.20 T^{-1/2}$ ns, with $T$ measured in Kelvin.  In the case of
short-pulse excitation, for $T=60$K, $\tau_*\approx$ 25 ps.  The present data
is consistent with the predicted approach to the adiabat of stability, but
since the photoluminescence time resolution in the experiments, from 100 to
300 ps, is much larger than $\tau_*$, the approach cannot be verified in
detail.

\section{Summary}

    We have shown that the observed quantum saturation of orthoexcitons is
caused by a competition between the Auger process and the cooling of the
excitons by the phonons.  When the orthoexcitons approach the phase boundary,
these two mechanisms act together to keep the orthoexcitons on a
particular adiabat, parallel, but at slightly higher temperature than the
phase boundary.  The cooling is affected by the proximity of the excitons to
the phase boundary, but the Auger process is not significantly affected by the
degeneracy of excitons.  We find that there is a fairly wide range of
numbers that can be inserted for the various parameters and still account for
the observed quantum saturation effect. Reasonable changes in the parameter
values cause the trajectories to shift relative to the phase boundary but
do not qualitatively change the results.  Also, the assumed laser profile
and width do not change the results of our calculation appreciably. We
have, however, made the assumption of a constant gas volume over the short
time intervals considered (1-20 ns).

    Our key finding is that, while the orthoexcitons encounter a formidable
barrier to condensation, the paraexcitons do indeed cross the phase boundary
and condense.  Specific features of the band structure of Cu$_2$O underlie
this condensation.  Earlier work has noted the simple parabolic bands and the
forbidden direct gap.  Our present study points out that
paraexciton-paraexciton Auger collisions are very slow for both the direct
and the phonon-assisted Auger processes.  Details will be reported in Ref.
\cite{auger}.  Finally we remark that this analysis presents no fundamental
reason why the orthoexcitons cannot condense; however their multiplicity,
their faster Auger decay and their conversion to paraexcitons make their
condensation much more difficult than for paraexcitons.

\acknowledgments

    The work of GMK and GB was supported by NSF Grants DMR91-22385 and
PHY94-21309, and that of JPW by NSF Grant DMR92-07458.  Helpful comments from
Y.C.  Chang, K. O'Hara, L. O'Suilleabhain and D.W.  Snoke are gratefully
acknowledged. GMK wishes to thank the Research Center of Crete, Greece for its
hospitality.

\vspace{1cm}
\begin{appendix}
\section{Calculation of electron (or hole)-exciton scattering matrix element}

    We calculate here the matrix element $M_{\vec q}^{ce}$ for scattering of
an electron or hole by an exciton (Fig.~9(b)). Consider an electron of momentum
$\vec k$ and an exciton of momentum $\vec p$ scattering into $\vec k - \vec q$
and $\vec p + \vec q$, respectively.  The electron (or hole)-exciton
interaction is:
\begin{eqnarray}
  {\cal{V}}= \frac {e^2} {\epsilon_0}
   \left( \frac 1 {|\vec r - \vec r_e|} - \frac 1 {|\vec r - \vec r_h|}
   \right),
\label{A1}
\end{eqnarray}
where $\vec r$ is the position of the electron (or hole) and $\vec r_e$
and $\vec r_h$ are the positions of the bound electron and hole, respectively.
The wavefunctions of the initial and final states are
\begin{eqnarray}
   \psi_i &=& \frac 1 {\sqrt V} e^{i \vec k \cdot \vec r}
             \Psi_{\vec p}(\vec r_e - \vec r_h), \nonumber\\
\label{A2}
   \psi_f &=& \frac 1 {\sqrt V} e^{i (\vec k - \vec q) \cdot \vec r}
   \Psi_{\vec p + \vec q}(\vec r_e - \vec r_h),
\label{A3}
\end{eqnarray}
where
\begin{eqnarray}
    \Psi_{\vec p} (\vec r_e - \vec r_h) = \frac 1 {\sqrt V}
     e^{i \vec p \cdot (\lambda_e \vec r_e +\lambda_h \vec r_h) }
     \Phi_{rel}(\vec r_e - \vec r_h)
\label{A4}
\end{eqnarray}
is the exciton wavefunction with center-of-mass momentum $\vec p$, and
$\Phi_{rel}(\vec r_e - \vec r_h)$ is the wavefunction of the relative motion.
The matrix element of the interaction between the initial and final states is
thus
\begin{eqnarray}
   M_{\vec q}^{ce} = \frac {4 \pi e^2} {\epsilon_0 V q^2}
   \left[\frac{1}{[1+(qa\lambda_h/2)^2]^2}
        -\frac{1}{[1+(qa\lambda_e/2)^2]^2} \right].
\label{A5}
\end{eqnarray}
The factor inside the parentheses comes again from the Fourier transform of the
relative electron-hole wavefunction, which we assume to be the 1s-state.
In the limit $\lambda_e \approx \lambda_h$,
\begin{eqnarray}
   M_{\vec q}^{ce} \approx \frac{2\pi e^2 a^2}{\epsilon_0 V}
   \frac {(\lambda_e-\lambda_h)}{\left(1+(qa/4)^2\right)^3}.
\label{A6}
\end{eqnarray}

\end{appendix}

    \figure{FIG. 1. (a)-(c) The LO phonon-assisted recombination
spectrum of orthoexcitons for three different times following a 100 ps laser
pulse.  The data are fitted by Bose-Einstein distributions (dashed lines).
Temperatures and densities extracted from the fits to the photoluminescence
spectra.  (d) Open circles show the density of orthoexcitons at various times
calculated from experimentally determined $\mu$ and $T$, as a function of
temperature, for short laser pulses.  Black dots are the results for
orthoexcitons in the long-pulse (10 ns) case.  In both experiments the
orthoexcitons move along the critical line for condensation, without
condensing.  All data is from Snoke et al.  Refs.  \cite{SWM87,SWM90} and
\cite{SW90}.}

    \figure{FIG. 2. (a) Data from lightly stressed crystals with long-pulse
(10 ns) excitation \cite{LW}.  The laser profile (triangles), the number of
orthoexcitons in the lowest orthoexciton level (open circles), and the number
of paraexcitons (black dots) as function of time.  The paraexcitons show a
significantly smaller decay rate.  (b) Corresponding trajectories for
orthoexcitons (open circles) and paraexcitons (black dots) in the
density-temperature plane.  The straight line is the condensation phase
boundary, which is identical for paraexcitons and orthoexcitons in the
stressed case.  Note that the paraexcitons are in the condensed region at
times later than 8 ns.}

    \figure{FIG. 3. Exciton-phonon interaction processes for non-condensed
(a,b) and condensed (c,d) excitons.}

    \figure{FIG. 4. Acoustic and optical phonon cooling rates per exciton as
a function of $\alpha$ for fixed exciton gas temperature 30K and lattice
temperature 6K.  Increased degeneracy (smaller $\alpha$) lowers the cooling
rate.}

    \figure{FIG. 5. Orthoexciton-paraexciton (a,b) down-conversion and
paraexciton-orthoexciton up-conversion (c,d) for acoustic phonon-mediated
mechanisms.}

    \figure{FIG. 6. The down-conversion and up-conversion rates are shown
for $\alpha_o=\alpha_p=0$ and $\alpha_o=\alpha_p=0.1$ as a function of the
temperature of the exciton gas.}

    \figure{FIG. 7. Direct Auger non-radiative annihilation processes.
Time progresses from left to right. The initial state contains two excitons
of momenta $\vec K$ and $\vec P$, and the final state contains an ionized
electron and hole with momenta $\vec k_e$ and $\vec k_h$, respectively.
The dashed line denotes the Coulomb interaction.}

    \figure{FIG. 8. Phonon-assisted Auger non-radiative annihilation processes.
The wiggly line denotes a phonon of momentum $\vec Q$.}

    \figure{FIG. 9. Energy transfer processes among hot electrons and holes
(single lines), excitons (double lines), and phonons (wiggly lines).  (a) and
(b) describe (Auger or laser generated) electron (hole)-phonon and electron
(hole)-exciton scattering processes, respectively; (c) hot exciton-thermalized
exciton scattering; (d) hot exciton-phonon scattering.}

    \figure{ FIG. 10. The ratio of the Auger heating $E_A$ over the binding
energy $E_b$ versus the density of the thermalized excitons, at fixed
exciton temperature, 30K.}

    \figure{FIG. 11. Numerical solution of Eqs.  (\ref{21}-\ref{23}) in the
density-temperature plane for the orthoexcitons (solid line) and paraexcitons
(dashed line), for a short laser pulse.  The diamonds are (long-pulse)
experimental data for orthoexcitons.  The two straight lines are Bose-Einstein
condensation boundaries for orthoexcitons (lower) and paraexcitons (higher)
The dash-dot line
shows the adabiat of stability (see Eq.  (\ref{s3})) with only
acoustic-phonon cooling and Auger heating taken into account.}

    \figure{FIG. 12. (a) Calculated behavior of orthoexcitons for three
different lattice temperatures, 18K (higher), 8K (middle) and 4K (lower),
assuming a short laser pulse.  (b) Behavior of orthoexcitons assuming
classical and degenerate statistics for the exciton-phonon interaction.}

    \figure{FIG. 13. Short laser-pulse excitation.  No ortho-para
interconversion process.}

    \figure{FIG. 14. Short laser-pulse excitation.  (a) Auger rate reduced
by 50\%; (b) Auger process turned off.}

    \figure{FIG. 15. The behavior of ortho and paraexcitons in the
density-temperature plane, for a long (10 ns) laser pulse. The dash-dot line
shows the adabiat of stability (see Eq.  (\ref{s3})) with only
acoustic-phonon cooling and Auger heating taken into account.}

    \figure{FIG. 16. The build-up (a) and decay (b) phase of Fig. 15.}

    \figure{FIG. 17. Behavior of ortho and paraexcitons assuming classical
statistics for the exciton-phonon interaction, for long laser-pulse
excitation.}

    \figure{FIG. 18. Behavior of ortho and paraexcitons in the
density-temperature plane, ignoring the interconversion process, for long
laser-pulse excitation.}

    \figure{FIG. 19. Behavior of ortho and paraexcitons in the
density-temperature plane, ignoring the Auger (heating and decay) process, for
long laser-pulse excitation.}

    \figure{FIG. 20. Behavior of ortho and paraexcitons in the
density-temperature plane, for long laser pulse excitation, assuming that the
scattering radius for exciton-exciton collisions (a) reduced by a factor of
ten, and (b) ten times larger than in Fig. 15.}

    \figure{FIG. 21. The shaded region shows the range of the values of the
Auger heating parameter $C$ and the magnitude of the acoustic-phonon
deformation potential $D_a$ which gives the observed quantum saturation of
the orthoexcitons.}

\end{document}